\documentclass{article}[11pt]
\usepackage{titling}
\usepackage{ amssymb }
\usepackage{amsfonts}
\usepackage{amsthm}
\usepackage{amsmath}
\usepackage{bbm}
\RequirePackage{etex}

\usepackage[utf8]{inputenc}
\renewcommand{\baselinestretch}{1.2}

\def\showauthornotes{0}

\def\showkeys{0}
\def\showdraftbox{0}
\def\showcolorlinks{1}
\def\usemicrotype{1}
\def\showfixme{0}

\usepackage{csquotes}

\usepackage{appendix}

\title{Locally testable codes via high-dimensional expanders\thanks{Part of this work was done when the authors were visiting the Simons Institute of Theory of Computing, Berkeley for the summer cluster on "Error-Correcting Codes and High-Dimensional Expansion".}}
\author{Yotam Dikstein\thanks{Weizmann Institute, {\tt yotam.dikstein@weizmann.ac.il}} \and Irit Dinur\thanks{Weizmann and IAS, {\tt irit.dinur@weizmann.ac.il}. Research supported by ERC CoG grant 772839 and the National Science Foundation under agreement No. CCF-1900460.} \and Prahladh Harsha\thanks{Tata Institute of Fundamental Research, Mumbai, India. {\tt prahladh@tifr.res.in}. Research supported by the Department of Atomic Energy,
    Government of India, under project no. 12-R\&D-TFR-5.01-0500 and in part by the Swarnajayanti fellowship.} \and Noga Ron-Zewi\thanks{Haifa University, {\tt noga@cs.haifa.ac.il}
}}
\date{\today}

\usepackage{etex}

\usepackage[l2tabu, orthodox]{nag}

\usepackage{xspace,enumerate}

\usepackage[dvipsnames]{xcolor}

\usepackage[T1]{fontenc}
\usepackage[full]{textcomp}

\usepackage[american]{babel}

\usepackage{mathtools}

\usepackage{amsthm}
\usepackage{amsmath}

\newtheorem{theorem}{Theorem}[section]
\newtheorem*{theorem*}{Theorem}

\newtheorem*{proposition*}{Proposition}
\newtheorem{lemma}[theorem]{Lemma}
\newtheorem*{lemma*}{Lemma}

\newtheorem*{corollary*}{Corollary}
\newtheorem*{conjecture*}{Conjecture}
\newtheorem{fact}[theorem]{Fact}
\newtheorem*{fact*}{Fact}

\newtheorem*{hypothesis*}{Hypothesis}

\theoremstyle{definition}
\newtheorem{definition}[theorem]{Definition}
\newtheorem*{definition*}{Definition}

\theoremstyle{remark}
\newtheorem{claim}[theorem]{Claim}
\newtheorem*{claim*}{Claim}
\newtheorem{remark}[theorem]{Remark}
\newtheorem*{remark*}{Remark}

\newtheorem*{observation*}{Observation}

\usepackage[letterpaper,
top=1in,
bottom=1in,
left=1in,
right=1in]{geometry}

\usepackage[varg]{pxfonts} % varg - uses nicer g,v,y,w

\usepackage{mathpazo}
\usepackage{lmodern}

\ifnum\showkeys=1
\usepackage[color]{showkeys}
\fi

\ifnum\showcolorlinks=1
\usepackage[
colorlinks=true,
urlcolor=blue,
linkcolor=blue,
citecolor=OliveGreen,
]{hyperref}
\fi

\ifnum\showcolorlinks=0
\usepackage[
colorlinks=false,
pdfborder={0 0 0}
]{hyperref}
\fi

\usepackage{prettyref}

\newcommand{\savehyperref}[2]{\texorpdfstring{\hyperref[#1]{#2}}{#2}}

\newrefformat{eq}{\savehyperref{#1}{\textup{(\ref*{#1})}}}
\newrefformat{lem}{\savehyperref{#1}{Lemma~\ref*{#1}}}
\newrefformat{def}{\savehyperref{#1}{Definition~\ref*{#1}}}
\newrefformat{thm}{\savehyperref{#1}{Theorem~\ref*{#1}}}
\newrefformat{cor}{\savehyperref{#1}{Corollary~\ref*{#1}}}
\newrefformat{cha}{\savehyperref{#1}{Chapter~\ref*{#1}}}
\newrefformat{sec}{\savehyperref{#1}{Section~\ref*{#1}}}
\newrefformat{app}{\savehyperref{#1}{Appendix~\ref*{#1}}}
\newrefformat{tab}{\savehyperref{#1}{Table~\ref*{#1}}}
\newrefformat{fig}{\savehyperref{#1}{Figure~\ref*{#1}}}
\newrefformat{hyp}{\savehyperref{#1}{Hypothesis~\ref*{#1}}}
\newrefformat{alg}{\savehyperref{#1}{Algorithm~\ref*{#1}}}
\newrefformat{rem}{\savehyperref{#1}{Remark~\ref*{#1}}}
\newrefformat{item}{\savehyperref{#1}{Item~\ref*{#1}}}
\newrefformat{step}{\savehyperref{#1}{step~\ref*{#1}}}
\newrefformat{conj}{\savehyperref{#1}{Conjecture~\ref*{#1}}}
\newrefformat{fact}{\savehyperref{#1}{Fact~\ref*{#1}}}
\newrefformat{prop}{\savehyperref{#1}{Proposition~\ref*{#1}}}
\newrefformat{prob}{\savehyperref{#1}{Problem~\ref*{#1}}}
\newrefformat{claim}{\savehyperref{#1}{Claim~\ref*{#1}}}
\newrefformat{relax}{\savehyperref{#1}{Relaxation~\ref*{#1}}}
\newrefformat{red}{\savehyperref{#1}{Reduction~\ref*{#1}}}
\newrefformat{part}{\savehyperref{#1}{Part~\ref*{#1}}}
\newrefformat{ex}{\savehyperref{#1}{Example~\ref*{#1}}}

\newcommand{\Sref}[1]{\hyperref[#1]{\S\ref*{#1}}}

\usepackage{nicefrac}
\ifnum\usemicrotype=1
\usepackage{microtype}
\fi

\ifnum\showauthornotes=1
\newcommand{\Authornote}[2]{{\sffamily\small\color{red}{[#1: #2]}}}
\newcommand{\Authornotecolored}[3]{{\sffamily\small\color{#1}{[#2: #3]}}}
\newcommand{\Authorcomment}[2]{{\sffamily\small\color{gray}{[#1: #2]}}}
\newcommand{\Authorstartcomment}[1]{\sffamily\small\color{gray}[#1: }

\newcommand{\Authorfnote}[2]{\footnote{\color{red}{#1: #2}}}
\newcommand{\Authorfixme}[1]{\Authornote{#1}{\textbf{??}}}
\newcommand{\Authormarginmark}[1]{\marginpar{\textcolor{red}{\fbox{\Large #1:!}}}}
\else
\newcommand{\Authornote}[2]{}
\newcommand{\Authornotecolored}[3]{}
\newcommand{\Authorcomment}[2]{}
\newcommand{\Authorstartcomment}[1]{}

\newcommand{\Authorfnote}[2]{}
\newcommand{\Authorfixme}[1]{}
\newcommand{\Authormarginmark}[1]{}
\fi

\ifnum\showfixme=0

\fi

\usepackage{boxedminipage}

\newcommand{\Brac}[1]{\left[#1\right]}

\newcommand\sett[2]{\left\{ #1 \left| \; \vphantom{#1 #2} \right. #2  \right\}}
\newcommand{\set}[1]{\{#1\}}

\newcommand{\Esymb}{\mathbb{E}}
\newcommand{\Psymb}{\mathbb{P}}

\DeclareMathOperator*{\E}{\Esymb}

\DeclareMathOperator*{\ProbOp}{\Psymb}

\renewcommand{\Pr}{\ProbOp}

\newcommand{\prob}[1]{\Pr \left[ {#1} \right] }
\newcommand{\Prob}[2][]{\Pr_{{#1}}\left[#2\right]} % use by \Prob[x]{event}
\newcommand{\cProb}[3]{\Pr_{{#1}}\left[ #2 \left| \; \vphantom{#2 #3} \right. #3  \right]} % use by \Prob[x]{event}

\newcommand{\Ex}[2][]{\E_{{#1}}\Brac{#2}}

 \usepackage{dsfont}
\usepackage{mathrsfs}

\newcommand{\textparen}[1]{\text{(#1)}}

\ifx\because\undefined
\newcommand{\because}[1]{\textparen{because #1}}
\else
\renewcommand{\because}[1]{\textparen{because #1}}
\fi

\newcommand{\bits}{\{0,1\}}

\newcommand{\defeq}{\stackrel{\mathrm{def}}=}

\newcommand\bdot\bullet

\DeclareMathOperator{\dist}{dist}

\newcommand{\etal}{et al.\xspace}

\renewcommand{\leq}{\leqslant}
\renewcommand{\le}{\leqslant}
\renewcommand{\geq}{\geqslant}

\ifnum\showdraftbox=1

\else

\fi

\let\epsilon=\varepsilon

\numberwithin{equation}{section}

\newcommand{\MYstore}[2]{%
  \global\expandafter \def \csname MYMEMORY #1 \endcsname{#2}%
}

\newcommand{\MYload}[1]{%
  \csname MYMEMORY #1 \endcsname%
}

\newcommand{\MYnewlabel}[1]{%
  \newcommand\MYcurrentlabel{#1}%
  \MYoldlabel{#1}%
}

\newcommand{\MYdummylabel}[1]{}

\newcommand{\torestate}[1]{%
  \let\MYoldlabel\label%
  \let\label\MYnewlabel%
  #1%
  \MYstore{\MYcurrentlabel}{#1}%
  \let\label\MYoldlabel%
}

\newcommand{\restatetheorem}[1]{%
  \let\MYoldlabel\label
  \let\label\MYdummylabel
  \begin{theorem*}[Restatement of \prettyref{#1}]
    \MYload{#1}
  \end{theorem*}
  \let\label\MYoldlabel
}

\newcommand{\restatelemma}[1]{%
  \let\MYoldlabel\label
  \let\label\MYdummylabel
  \begin{lemma*}[Restatement of \prettyref{#1}]
    \MYload{#1}
  \end{lemma*}
  \let\label\MYoldlabel
}

\newcommand{\restateprop}[1]{%
  \let\MYoldlabel\label
  \let\label\MYdummylabel
  \begin{proposition*}[Restatement of \prettyref{#1}]
    \MYload{#1}
  \end{proposition*}
  \let\label\MYoldlabel
}
\newcommand{\restatefact}[1]{%
  \let\MYoldlabel\label
  \let\label\MYdummylabel
  \begin{fact*}[Restatement of \prettyref{#1}]
    \MYload{#1}
  \end{fact*}
  \let\label\MYoldlabel
}

\newcommand{\restateclaim}[1]{%
  \let\MYoldlabel\label
  \let\label\MYdummylabel
  \begin{claim*}[Restatement of \prettyref{#1}]
    \MYload{#1}
  \end{claim*}
  \let\label\MYoldlabel
}

\newcommand{\restatecorollary}[1]{%
  \let\MYoldlabel\label
  \let\label\MYdummylabel
  \begin{corollary*}[Restatement of \prettyref{#1}]
    \MYload{#1}
  \end{corollary*}
  \let\label\MYoldlabel
}

\newcommand{\restatedefinition}[1]{% % overwrite label command with dummy
\let\MYoldlabel\label 
\let\label\MYdummylabel 
\begin{definition*}[Restatement of \prettyref{#1}] 
    \MYload{#1} 
\end{definition*} 
\let\label\MYoldlabel 
} 

\newcommand{\restate}[1]{%
  \let\MYoldlabel\label
  \let\label\MYdummylabel
  \MYload{#1}
  \let\label\MYoldlabel
}

\newcommand{\eps}{\epsilon}

\let\origparagraph\paragraph
\renewcommand{\paragraph}[1]{\origparagraph{#1.}}

\allowdisplaybreaks

\let\pref=\prettyref

\renewcommand{\restriction}{|}%\mathord{\upharpoonright}}
\newcommand{\rest}[2]{{#1}\restriction_{{#2}}}

\newcommand{\bigO}[1]{O \left( #1 \right)}

\begin{document}
\maketitle

\def\c{\textbf{c}}
\def\A{{\mathcal{A}}}
\begin{abstract}

Locally testable codes (LTC) are error-correcting codes that have a local tester which can distinguish valid codewords from words that are far from all codewords, by probing a given word only at a very small (sublinear, typically constant) number of locations. Such codes form the combinatorial backbone of PCPs. A major open problem is whether there exist LTCs with positive rate, constant relative distance and testable with a constant number of queries. 

In this paper, we present a new approach towards constructing such LTCs using the machinery of high-dimensional expanders. 
To this end, we consider the Tanner representation of a code, which is specified by a graph and a base code. Informally, our result states that if this graph is part of an {\em agreement expander} then the local testability of the code follows from the local testability of the base code. Agreement expanders allow one to stitch together many mostly-consistent local functions into a single global function. High-dimensional expanders are known to yield agreement expanders with constant degree. 

This work unifies and generalizes the known results on testability of the Hadamard, Reed-Muller and lifted codes, all of which are proved via a single round of local self-correction: the corrected value at a vertex $v$ depends on the values of all vertices that share a constraint with $v$. In the above codes this set includes all of the vertices. In contrast, in our setting the degree of a vertex might be a constant, so we cannot hope for one-round self-correction. We overcome this technical hurdle by performing iterative self correction with logarithmically many rounds and tightly controlling the error in each iteration using properties of the agreement expander.

Given this result, the missing ingredient towards constructing a constant-query LTC with positive rate and constant relative distance is an instantiation of a base code and a constant-degree agreement expander that interact well with each other.
\end{abstract}

\section{Introduction} \label{sec:intro}

In this work, we study an approach to constructing locally testable codes (LTCs) based on high-dimensional expansion. LTCs are error-correcting codes that have a local tester which can test if a given word is a valid codeword or far (in Hamming distance) from all codewords, by probing the given word only at a very small (sublinear, typically constant) number of locations. 
Reed-Muller codes were the first codes shown to be locally-testable \cite{FriedlS1995,RubinfeldS1996}. These codes are based on low degree polynomial functions, and have inverse polynomial rate. Later on, LTCs with inverse poly-logarithmic rate were constructed by \cite{BenSassonS2008,Dinur2007}. Obtaining an LTC family with rate that is not vanishing is a major open question in this area. Such codes are known as ``good'' LTCs or $c^3$-LTCs since they have \c onstant rate, \c onstant relative distance, and testable with a \c onstant number of queries~\cite{Goldreich2010}. This question is interesting in its own right, and also could potentially lead towards constructing linear-length PCPs (as LTCs are the combinatorial backbone of all PCP constructions). 
The problem of constructing $c^3$-LTCs is particularly difficult as we do not know if such good codes exist, even non-explicitly (say using a probabilistic argument). The difficulty stems from the fact that local testability requires redundancy in the constraints. In known LTCs, the constraints are highly overlapping, a property that in the past went hand in hand with relatively {dense} families of constraints. Alas this density seems to significantly limit the rate. In contrast, high-dimensional expanders give {sparse} families of subsets that are heavily overlapping. Perhaps if we manage to find appropriate constraints on these subsets we may find higher rate LTCs. 

In this work, the vague notion of ``overlapping constraints'' is captured through so-called agreement-expansion (which will be formally defined below). 

Informally speaking, we show that if an error-correcting code is defined through a collection of local constraints that {\em sit on an agreement expander}, then to prove local testability of the entire code it suffices to prove local testability of the local components (which are of merely constant size in the case of constant-degree agreement expanders). This is similar in spirit to recent applications of high-dimensional expanders towards proving other local-to-global results. This passing from local to global is particularly important because known constructions of high-dimensional expanders are very difficult to analyze on a global level. So far, successful analyses focused on the local structure (in neighborhoods, or so-called links) of these objects. Through this work, the task of constructing global LTCs is reduced to the task of constructing LTCs on the local structure, which appears to be a much more reasonable task. 

This work can be viewed as providing a generic scheme for constructing an LTC on a high-dimensional expander (or an agreement expander), and the (big) missing ingredient is an appropriate instantiation. We comment that the flagship example of an LTC, namely Reed-Muller codes, can be viewed as an instantiation of this scheme, with the underlying agreement expander being the Grassmannian complex and the base code being the Reed-Solomon code (see \pref{sec:applications}). The hope is that replacing the ``dense'' Grassmannian complex by a bounded-degree complex, together with finding an appropriate base code, could potentially lead to a $c^3$ LTC.

\paragraph{Tanner Codes} To elucidate the main result, we begin by recalling a well-studied family of codes, the \emph{Tanner codes}~\cite{Gallager1960,Tanner1981}. A Tanner code $C \subseteq \bits^n$ is given by a family of (small, often constant-sized) subsets $t_1,\ldots,t_m \subset [n]$ and for each subset a base code $C_{t_i}\subset \bits^{t_i}$. A string $w\in \bits^n$ is in the code $C$ if for each $i$, $\rest{w}{t_i}\in C_{t_i}$.\footnote{A Tanner code is equivalently described on a bipartite graph (called the Tanner graph) with $n$ right vertices corresponding to the coordinates of the code and $m$ left vertices corresponding to the sets $t_i$, with an edge between $v$ and $t_i$ if $v\in t_i$. } Many known codes, including Reed-Muller codes, lifted codes, tensor codes, and expander codes, are in fact Tanner codes. In all of these cases, there is a single base code $C_0$ such that $C_{t_i} = C_0$ for all $i$, but this need not be the case. 

The Tanner representation of a code also gives a natural candidate for a local test for checking whether a given word $w\in \bits^n$ is in the code.

\textbf{Natural Tanner Test}: Choose a random $i\in [m]$ and accept iff $\rest{w}{t_i}\in C_{t_i}$. 

We say that $C$ is $\rho$-locally-testable with the natural tester if
\[ \rho\cdot \dist(w,C) \le  \prob{\text{Test fails}}.\] 
A family of codes is a locally testable code (LTC) if it satisfies the above inequality for some test (not necessarily the natural Tanner test) with a constant $\rho$ (that does not decrease with the block length of the code).

Many Tanner codes, including expander codes and random LDPC codes, that are very good in terms of rate and distance, {\em and} can be characterized by ``low density'' constraints (that look at only a constant number of bits in the codeword) fail quite miserably at being LTCs~\cite{BenSassonHR2005}. \\

Imagine that in addition to $T=\set{t_1,\ldots,t_m}$ we also have a family $S$ of subsets of $[n]$, such that each $s\in S$ has constant size, but slightly larger than the size of the $t_i$'s. For each such $s\in S$ we consider the `local' Tanner code \[C_s = \sett{w\in \bits^s}{w|_t \in C_t,\,\forall t\in T,\;t\subset s}.\]
(Of course, $C_s$ is non-trivial only if there are some $t\in T$ contained in $s$.)

In this work, we show that if for each $s\in S$, the code $C_{s}$ itself is locally testable with the natural Tanner test, then the code $C$ too must be locally testable with respect to the natural Tanner test. This holds as long as we assume some nice structure on the families $S$ and $T$, namely that they are part of a ``multi-layered agreement sampler'', MAS for short, which is described below.  

Let us change point of view and look at the codes $\set{C_s}$ as a collection of base codes, giving rise to the Tanner code $C$. Our main result is that local testability of the base codes $C_s$ \emph{lifts} to local testability of the entire code $C$, assuming an expander-like MAS condition on the underlying Tanner graph. This is analogous to the celebrated expander codes \cite{SipserS1996} in which distance of the base codes gets lifted to distance of the entire code, assuming expansion of the underlying Tanner graph. Whereas expansion alone does not suffice for local testability, the MAS structure does.
\paragraph{High-dimensional expanders and Agreement Expanders}
There are several interesting and non-equivalent definitions for high-dimensional expanders, the two main ones being topological definitions of coboundary or cosystolic expansion \cite{LinialM2006,Gromov2010,DotterrerKW2018}, and, more relevant to this work, random walk definitions either locally at the link level \cite{KaufmanM2017, DinurK2017} or globally \cite{DiksteinDFH2018, KaufmanO2017}.
Without going into details, high-dimensional expansion has already been shown to imply some surprising local to global theorems. For example the trickling down theorem of \cite{Oppenheim2018} proves global spectral expansion using local spectral expansion in the links (which are the neighborhoods of individual vertices). Another example is the list decoding of \cite{DinurHKNT2019} which deduces global list decoding from list-decoding on the local pieces.

Yet another example, which is crucial for this work, is that high-dimensional expanders give rise to agreement expanders~\cite{DinurK2017,DiksteinD2019}. An agreement expander allows one to stitch together many mostly-consistent local functions into a single global function. We elaborate a little more on this notion. Let $V$ be a ground set of $n$ elements, and let $
S$ be a collection of subsets of $V$ of some fixed size. Let $\A$ be a graph whose vertices are the subsets in $S$, and each edge $\set{s,s'}$ is labeled by a subset $k\subset s\cap s'$. Let $K$ be a collection of subsets labelling the edges. 

$(V,K,S,\A)$ is an $\alpha$-agreement expander if whenever an ensemble has agreement value $1-\eps$ there exists a global function $F\colon V \to \{0,1\}$ such that $f_s = F|_s$ for all but at most $\eps/\alpha$ of $s\in S$. (See \pref{sec:ae} for the full definition).
An agreement expander is given by $V,K,S$ and the edge-labelled graph $\A$. Suppose that for each $s\in S$ we are given a local function $f_s\in \bits^s$. The {\em agreement value} of the ensemble $\set{f_s}$ is the probability of $f_s|_k = f_{s'}|_k$ for a randomly chosen edge $\set{s,s'}_k$ (this is notation for an edge between $s,s'$ labeled by $k$) in the graph $\A$. Whenever there is a global function $F\colon V\to\bits$ such that $f_s = F|_s$ for all $s\in S$, the agreement value of $\set{f_s}$ is clearly $1$. We say that

Agreement expanders have been studied and used in the LTC and PCP literature for years (under different names such as direct product tests or sometimes low degree tests). However, prior to the recent connection with high-dimensional expanders, the only known agreement expanders were relatively dense. The existence of sparse such objects seems promising and could potentially lead to LTCs with positive rate. This work shows how agreement expansion can be useful for constructing LTCs. 

\paragraph{Multilayered Agreement Samplers (MAS)} 
We now describe the MAS combinatorial structure needed for our LTC scheme.
Let $V$ be a ground set of $n$ elements, and let $T,K,S$ be three families of subsets of $V$ of sizes $q_0<q_1<q_2$.
The system $(V,T,K,S)$ is said to be a \((\lambda,\alpha)\) -\emph{MAS} if the following two conditions are met.
\begin{itemize}
    \item $V,K,S$ are part of an $\alpha$-agreement-expander. 
    \item The bipartite containment graph of \(T\) vs. \(K\) is a \(\lambda\)-sampler. 
\end{itemize}
The above definition is stricter than what we actually need, see the formal more refined definition in \pref{def:mas}. We are now ready to state our main result.

\paragraph{Main Result} 
Let $V,T,K,S$ be a \((\lambda,\alpha)\) -\emph{MAS}. 
Suppose that for each $t\in T$ we have a local code \(C_t\subset \bits^t\). 
Let $C \subset \bits^n$ be the Tanner code defined by $\set{C_{t}}$ for all $t\in T$. Namely,
\[
C:= \sett{w\in \bits^V}{\rest{w}{t}\in C_t\hbox{ for every }t}.\]  
Similarly, for each $s\in S$, let $C_s$ be the Tanner code defined by $\sett{C_t}{t\subset s}$, namely,
\[
C_s = \sett{w\in \bits^s}{\rest{w}{t}\in C_t\hbox{ for every }t\subset s},\] 
and similarly define for each $k\in K$, \(C_k = \sett{w\in \bits^k}{\rest{w}{t}\in C_t\hbox{ for every }t\subset k}\).

\begin{theorem}\torestate{ \label{thm:main-hdx}
Let $V,T,S$ be layers in a $(\lambda,\alpha)$-MAS satisfying $\lambda \leq \rho \delta \alpha/64$.
Suppose $C_k\subset \bits^k$ has relative distance $\delta$ for all $k\in K$ and suppose that $C_s$ is $\rho$-locally testable with the natural Tanner tester. Then \(C\) is $\rho \delta \alpha/16$ locally testable (with the natural Tanner tester).} 
\end{theorem}
We state our full main theorem in \pref{thm:LLTC-imples-GLTC}.

\paragraph{Overview of proof} Our proof of local testability, like previous proofs of testability, goes via self correction. The main difficulty in our setting is that a single round of self-correction is insufficient to correct the word.

Let \(w\) be a word that satisfies a \((1-\varepsilon)\)-fraction of the constraints in the Tanner graph. We would like to show that there exists a \(w^* \in C\) such that \(\dist(w,w^*) = O(\varepsilon)\). For specific codes, one could use the properties of the code to perform this self-correction (cf. Reed-Muller testing, one could use the properties of polynomials).

However, we cannot resort to such properties since we are working in an abstract setting. Instead, we rely on simple majority decoding: each vertex takes a value that satisfies the majority of the constraints it participates in. The main engine driving our proof is agreement expansion. 
Our proof strategy is as follows:

Construct a word \(w'\) from the received word \(w\) via self correction (or otherwise) and show 
\begin{enumerate}
    \item[(a)] \(w\) is close to \(w'\), and
    \item[(b)] \(w'\) is a valid codeword.
\end{enumerate}
Property (a) is easy to show if \(w'\) is constructed via self correction using majority decoding. Property (b) is not very hard in the context of Hadamard testing and Reed-Muller testing: every vertex participates in a constraint with every other vertex (indeed the diameter of the Tanner graph is a constant), hence one round of self-correction results in a valid codeword $w'$. However, since our proof is general enough to work even for constant-degree Tanner graphs wherein the diameter can be as large as logarithmic, one does not expect a single step of self correction via majority decoding to yield a codeword in a single step. \

Our proof instead relies on a novel iterative self correction procedure that slowly corrects a given word in logarithmically many iterations. A standard problem that arises when using iterative procedures is that the error grows linearly in the number of iterations, which is prohibitively expensive in our setting. We use the properties of MAS to show that the number of unsatisfied constraints by the self-corrected word $w'$ reduces by a constant factor in each iteration. This allows us to perform an arbitrary number of rounds in the iterative self-correction procedure till we reach a perfect codeword $w^* \in C$ (actually a logarithmic number of rounds will suffice). This type of argument is new in the context of locally testable codes. 

Given this we can proceed with the proof overview as follows. Since \(w\) satisfies \((1- \varepsilon)\)-fraction of the constraints, an averaging argument shows that a \((1-O(\varepsilon))\)-fraction of the \(s\)'s satisfy most of the constraints within them. Hence, by the local testability of the code $C_s$ we get that for most \(s\)'s, \(\rest{w}{s}\) is close to a local codeword, say \(w_s\in C_s\). Furthermore, it is not hard to show that these local codewords
satisfy that for a typical $k \in K$ and \(s,s' \in S\) such that \(k \subset s\cap s'\), we have \(\rest{w_s}{k} \equiv \rest{w_{s'}}{k}\). In other words, the \(w_s\)'s satisfy the hypothesis of the agreement test. From the agreement expansion of the MAS, there exists a ``global'' word \(w'\) that explains most of the \(w_s\)'s. Furthermore, it is not hard to show that \(w'\) is close to the original word \(w\). We then use the sampler property of the MAS to show that \(w'\) violates {\em significantly fewer} constraints than \(w\) (in particular, \(w'\) violates at most \(\varepsilon/2\)-fraction of constraints).

We iteratively apply the above self-correction procedure to get a sequence of words such that \(w^{(0)} := w,w^{(1)},w^{(2)},\ldots\) such that
\(w^{(i)}\) violates at most \(\varepsilon/2^i\)-fraction of constraints and \(\dist(w^{(i)},w^{(i+1)}) = \bigO{\varepsilon}/2^i\). Since the fraction of violated constraints cannot infinitely decrease, we have that eventually for a large enough \(i\), \(w^*:=w^{(i)} \in C\) and \( \dist(w,w^*) \leq \sum_{j=0}^{i-1}dist(w^{(j)},w^{(j+1)}) = O(\varepsilon)\).

\paragraph{Relation to previous work} We begin by recalling the history of LTCs and the close connection between PCP and LTC constructions. LTCs were first studied in the context of program checking
by Blum, Luby and Rubinfeld~\cite{BlumLR1993} and Gemmell~{\em
  et al.}~\cite{GemmellLRSW1991}. The notion of LTCs is implicit in the work on locally
checkable proofs by Babai~\etal~\cite{BabaiFLS1991} and
subsequent works on PCPs. The explicit definition appeared
independently in the works of Rubinfeld and
Sudan~\cite{RubinfeldS1996}, Friedl and Sudan~\cite{FriedlS1995}, Arora's
PhD thesis~\cite{Arora1994} and Spielman's PhD thesis~\cite{Spielman1995}. A formal
study of LTCs was initiated by  Goldreich and
Sudan~\cite{GoldreichS2006}.   Most known constructions of PCPs yield LTCs with similar
parameters. In fact, there is a generic transformation to convert a
PCP of proximity (which is a PCP with more requirements) into
an LTC with comparable
parameters~\cite{BenSassonGHSV2006,Trevisan2004}. See a
survey by Goldreich~\cite{Goldreich2010} for the interplay between PCP
and LTC constructions. In fact, the current best construction of LTCs (constant-query, constant fractional distance and inverse polylogarithmic rate) is obtained from the PCP constructions of Ben-Sasson and Sudan~\cite{BenSassonS2008} and Dinur~\cite{Dinur2007}. PCP-based constructions are unlikely to yield LTCs with constant rate since PCP constructions typically involve at least a logarithmic overhead. Nevertheless LTC constructions that aren't derived from PCPs perhaps have a better chance at achieving the coding-theory gold-standard of positive rate and distance.

Agreement expansion and the multilayered set system structure play a central role in our proof of local testability. 
Another application of agreement expansion towards local testability was studied in \cite{DinurHKR2019}, where it was used to enhance the local testability of a code in the context of the subspaces (Grassmannian) complex.
We remark that use of such multilayered agreement samplers in the context of locally-testable codes is actually implicit in many previous constructions of locally testable codes. The Raz-Safra \cite{RazS1997} proof of the local testability of the Reed-Muller codes works with points-lines-planes structure, a subgraph of the Grassmannian complex which is an excellent agreement expander as explained in detail in \pref{sec:applications}. The original proof due to Blum, Luby and Rubinfeld \cite{BlumLR1993} (as well as subsequent improvements due to Coppersmith) of the local testability of the Hadamard codes as well as Kaufman and Sudan's proof of testability of affine-invariant codes~\cite{KaufmanS2008}, relies on the three-layered structure comprising of the points, the three-point tests and certain nine-point sets, sometimes referred to as "magic squares"~\cite{KaufmanS2008}.

Our proof makes explicit this use of MAS to construct LTCs and shows that four-layered MAS are sufficient to transform ``local'' local testability to ``global'' local testability. In this sense, our proof can be viewed as bringing together these seemingly different proofs of local-testability under a common umbrella.

We already remarked that our construction has a similar paradigm as the Sipser-Spielman construction of expander codes~\cite{SipserS1996} which demonstrates that if the base code has good distance then the Tanner code also has good distance provided the graph is an expander. Another construction of the same flavor is the result of Dinur et al.~\cite{DinurHKNT2019} that demonstrates that if the local code is efficiently list-decodable then so is the global code defined by ABNNR distance amplification property via an expander~\cite{AlonBNNR1992}, provided the expander is part of a large high-dimensional expander.

\paragraph{Further Discussion and Future Work} 

This work gives a general scheme for constructing an LTC. It needs to be instanciated with an appropriate MAS and base codes. As mentioned earlier, and explained in detail in \pref{sec:applications}, one such instanciation is to choose the Grassmannian complex as the MAS, and the Reed Solomon code as the base codes. This gives the well-studied locally testable codes called Reed-Muller codes, as well as the more recent so-called lifted codes. 

The most interesting direction is to instantiate this scheme with an MAS that comes from some bounded-degree high-dimensional expander, and to combine it with appropriate choice of locally testable base code. The main hurdle in choosing the base codes is to be able to certify that the resulting Tanner code maintains positive rate. In some similar situations this is done by a simple counting of the number of constraints. However, such an argument cannot work in the setting of LTCs, and we leave it as an open question.

\section{Preliminaries}
\label{sec:preliminaries}
\subsection{Error Correcting Codes}
Let \(\Sigma\) be some finite set. A code is some \(C \subseteq \Sigma^n\). Let \(p\) be a prime power and \(\Sigma = \mathbb{F}_p^n\) be an \(n\)-dimensional vector space over a field with \(q\) elements. We say that \(C\) is a \emph{linear code} when \(C\) is a subspace of \(\mathbb{F}_p^n\). The rate of the code is \(rate(C) = \frac{\log_q |C|}{n}\).

It is convenient to think about \(\mathbb{F}_p^n\) as functions \(f:[n] \to \mathbb{F}_p\). The distance between two functions \(f,g:[n] \to \Sigma\), denoted by \(\dist(f,g)\), is the fraction of \(x\in [n]\) so that \(f(x)\ne g(x)\). The distance of a code is defined to be \(\dist(C) = \min_{f,g\in C, f\ne g} \dist(f,g)\). When \(C\) is linear, this is the same as \(\min_{0\ne f\in C}\dist(f,0)\).

\subsection{Tanner Codes}
A Tanner code \cite{Gallager1960,Tanner1981} over an alphabet $\Sigma$ (also called a lifted code) is defined through two objects: a family \(T\) of \(q\)-element subsets of \([n]\), and with each subset $t\in T$ a base code \(C_t \subset \Sigma^t\). The code \(C \subseteq \Sigma^n\) is given by
\[ C = \sett{w \in \Sigma^n}{\rest{w}{t} \in C_t, t \in T}.\]
The family \(T\) is often described through a bipartite graph on vertex sets \([n]\) and \(T\) connecting \(t \in T\) to \(i \in [n]\) whenever \(i \in t\). Several well-known families of codes can be constructed as Tanner codes, including tensor codes, Reed-Muller codes, and the codes considered by Sipser and Spielman~\cite{SipserS1996}. A family of Tanner codes that is especially related to our context is the family of so-called lifted codes. Lifted codes were first introduced by Ben-Sasson, Maatouk, Shpilka and Sudan~\cite{BenSassonMSS2011} and their local testability was studied by Guo, Kopparty and Sudan~\cite{GuoKS2013}. These codes can be described as Tanner codes where \([n]\) is identified with points of a vector space and the family \(T\) contains all possible affine subspaces of a prescribed dimension \(m\). The base code \(C_0\) is taken to be affine invariant. A prime example for such codes is the Reed-Muller code. 

\subsection{Locally Testable Codes}
A \((Q,\rho)\)-local tester for the code \(C\) is a probabilistic oracle algorithm that determines whether a word is in the code. It does the following: given oracle access to a function \(f:[n] \to \Sigma\), it queries \(f\) at \(Q\) input locations. Then if \(f \in C\) it accepts with probability \(1\). If \(f \notin C\) it rejects with probability at least \(\rho \cdot dist(f,C)\). Here \(\rho \in (0,1)\) is some constant parameter, and \(dist(f,C)\) is the distance between \(f\) and the closest codeword to it in \(C\).

For linear codes \(C\), \cite{BenSassonHR2005} showed that without loss of generality, we can assume that the local testers picks a random subset \(t \subset [n]\) according to some distribution, and accept if and only if \(\rest{w}{t} \in C_t\) (that is, that there exists some codeword \(w' \in C\) so that \(\rest{w}{t} = \rest{w'}{t}\)). Thus we formally define the locally testable codes as following:

\begin{definition}[Locally Testable Codes]
Let \(V\) some finite set, and \(C\) be some linear code on \(V\). Let \(D\) be some distribution on subsets of \(V\), and suppose every set \(t \sim D\) of of size at most \(Q\). Let \(\rho > 0\). We say \(C\) is \((Q,\rho)\)-testable with respect to \(D\) if
\[ \rho \cdot \dist(f,C) \leq \Prob[t \sim D]{\rest{f}{t} \notin \rest{C}{t}}.\]
\end{definition}

An alternate way of describing a locally testable code is using the Tanner graph \(G= ([n],T,E)\) representation of a code. In this representation, \([n]\) corresponds to the \(n\)
input locations in the codeword. \(T\) corresponds to the subsets of indexes that are queried by the local tester. We connect \(i \in [n]\) and \(t\in T\) if \(i \in t\).

A local tester that corresponds to this representation picks a random constraint \(t \in T\) and checks if the corresponding constraint is satisfied.

\subsection{Sampler Graphs}
Let \(G = (U,V,E)\) be a bipartite graph,
and assume that each edge carries a non-negative weight \(p_{uv}\) such that \(\sum_{uv \in E} p_{uv} = 1\).
This probability distribution induces a marginal probability distribution on U and similarly on V given by \(p_u =\sum_{uv\in E}p_{uv}\).
For every set \(B \subseteq U\) (and \(V\) respectively) we denote by \(\prob{B} = \Prob[u \in U]{u \in B}.\)
As a slight abuse of notation, for a set \(B \subseteq U\) and a vertex \(v_0 \in V\) we denote by 
\[\cProb{}{B}{v_0} = \cProb{uv \in E}{u \in B}{v=v_0}.\]

A sampler graph is a graph where for all \(B \subseteq U\), most of the vertices \(v_0 \in V\) have that \(\prob{B} \approx \cProb{}{B}{v_0}\).
\begin{definition}[\(\lambda\)-sampler]
Let \(G = (V,U,E)\) be a bipartite graph. For any \(B \subset U\), we define
\(N = N(B,\delta) = \sett{v \in V}{\cProb{u \in U}{u \in B}{u \sim v} > \prob{B} + \delta}\). For \(\lambda \in (0,1)\) we say \(G\) is a \(\lambda\)-sampler if for every \(B \subseteq U\) and every \(\delta > 0\),
\[\prob{N} \leq \frac{\lambda}{\delta^2}\prob{B}.\]
\end{definition}

There is a tight connection between expander bipartite graphs and sampler graphs. For more on this, see \cite{Goldreich2011-samp}.

\subsection{Agreement Expanders}\label{sec:ae}
\def\A{{\mathcal{A}}}
Let \(V\) be a finite universe, $S$ a collection of subsets of $V$, and for each subset $s\in S$, a local function $f_s\in \Sigma^s$. An ensemble $\set{f_s}$ is \emph{perfectly global} if it comes from a single global function  $w:V\to \Sigma$, namely, \(f_s = \rest{w}{s}\) for all $s$. We denote by $\mathcal{G}$ the collection of all perfectly global ensembles.
An agreement tester is given by a non-negatively weighted graph $\A$ with vertex set $S$, and such that each edge $\set{s,s'}$ is labelled by some $k\subseteq s\cap s'$. Without loss of generality we require that the weights sum to $1$, so that the edges form a distribution over pairs $s,s'$. Given a collection $\set{f_s}$ of local functions, the tester selects an edge $s,s'$ at random, and accepts if $f_s(v) = f_{s'}(v)$ for each $v\in k$. We call this {\em the value of $\set{f_s}$ under $\A$} and denote it by  $\A(\set{f_s})$,
\[ \A(\set{f_s}) := \Pr_{s,s'\sim \A} [ f_s(v) = f_{s'}(v),\;\forall v\in k].
\]
It is clear that a perfectly global ensemble has value $1$. Indeed for any pair $s,s'$ and any $v'\in s\cap s'$, $f_s(v) = g(v) = f_{s'}(v)$ assuming that $g:V\to\Sigma$ is the global function that agrees with $\set{f_s}$. The graph $\A$ is an agreement expander if a robust converse holds, namely any ensemble with $A(\set{f_s})\approx 1$ has to be close to a perfectly global ensemble. Formally,
\begin{definition}
Let $V,K,S,\A$ be as above. We call $\A$ an $\alpha$-agreement expander if for every ensemble of local functions $\set{f_s}$
\begin{equation}\label{eq:c-soundness}
    \alpha \cdot \dist(\set{f_s},\mathcal{G}) \leq 1-\A(\set{f_s}).
\end{equation}
where the distance \(\dist(\set{f_s},\mathcal{G})\) is the distance between $\set{f_s}$ and the closest perfectly global ensemble; where distance between two ensembles $\set{f_s},\set{g_s}$ is defined as probability $f_s\neq g_s$ when $s$ is chosen from the marginal distribution of $\A$. 
\end{definition}

\paragraph{More refined notions of agreement-expansion}
We also say that {\em $\A$ is an agreement expander with respect to $\delta$-ensembles} if \eqref{eq:c-soundness} holds for every ensemble $\set{f_s}$ that is a \(\delta\)-ensemble, namely, such that for every edge $\set{s,s'}_k$ in the graph $\A$, we have either \(\rest{f_s}{k} = \rest{f_{s'}}{k}\) or else the Hamming distance between $\rest{f_s}{k}$ and $\rest{f_{s'}}{k}$ is at least \( \delta|k|\).

Furthermore, we allow a slightly weaker notion of distance from being perfectly global. We say that $\A$ has $(K,\alpha)$ soundness wrt $\delta$ ensembles if the following holds. Suppose that for every \(s \in S\) there is a distribution \(k \sim D_s\) that samples \(k \in K\) that are subsets of \(s\). We say that $\A$ is \((K,\alpha)\)-sound wrt $\set{f_s}$ when
\begin{equation}\label{eq:k-c-soundness}
    \alpha \cdot \min_{G \in \mathcal{G}}\Prob[s \in S, k\sim D_s]{\rest{f_s}{k} \ne \rest{G}{k}} \leq 1-\A(\set{f_s}).
\end{equation}
We say that $\A$ is $(K,\alpha)$ sound with respect to $\delta$ ensembles if \eqref{eq:k-c-soundness} holds for all $\delta$ ensembles. 

\section{Multilayer Agreement Samplers}
\label{sec:MAS}

\begin{figure}
    \centering
    \includegraphics[scale=0.4]{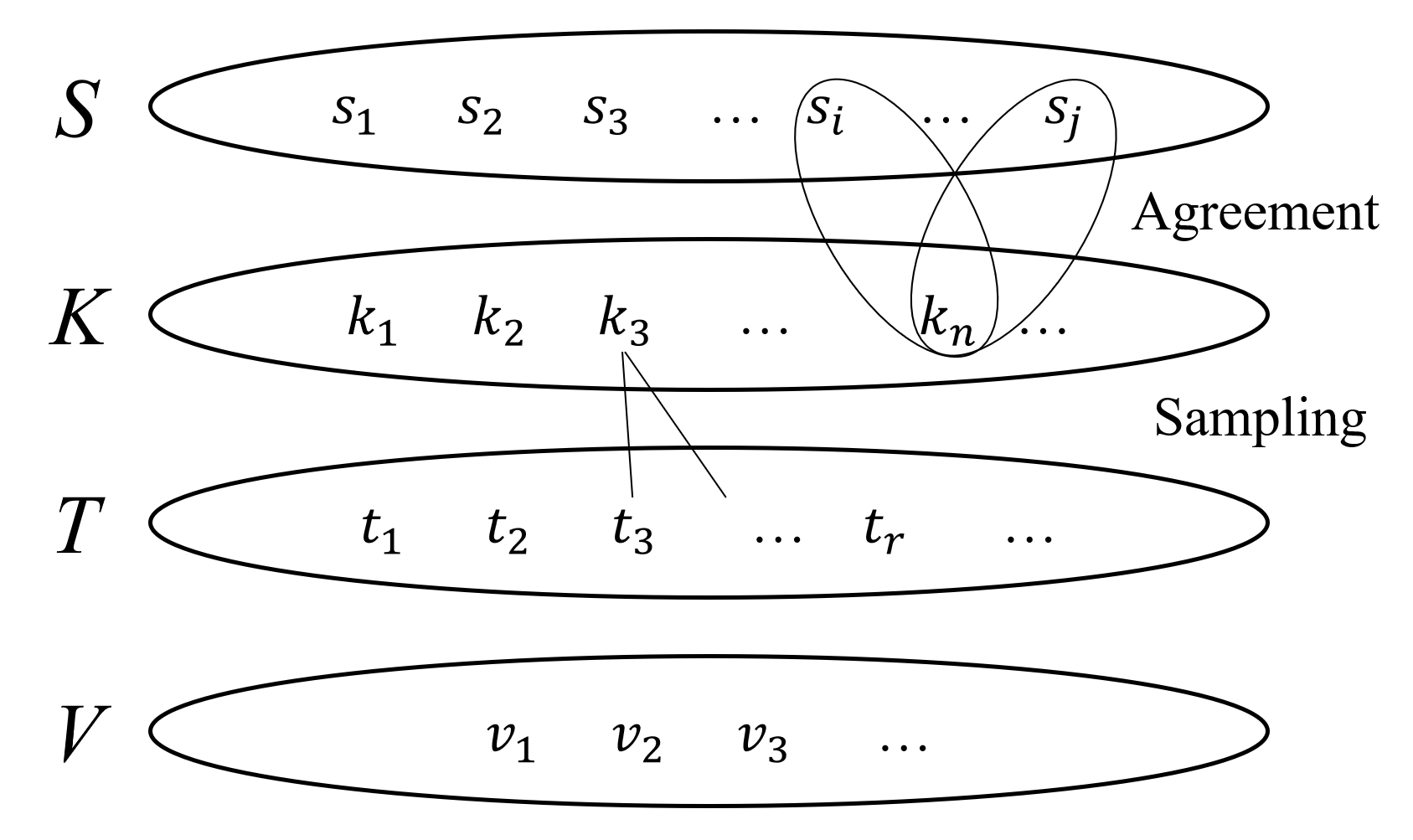}
    \caption{Multilayer Agreement Sampler}
    \label{fig:mas}
\end{figure}

The structure we use to construct locally testable codes has a sampler component and an agreement expander component, that sit together in four layers. We call these structures Multilayer Agreement Samplers.

\begin{definition}[Multilayer Agreement Samplers (MAS)]\label{def:mas}
Let \(\delta,\lambda,\alpha \geq 0\). Let \(V\) be a set of elements, and \(T,K,S \subset 2^{V}\) be families of subsets so that there is a non-degenerate Markov chain that samples \((v,t,k,s)\) from \(V,T,K,S\) respectively, so that \(v \in t \subset k \subset s\). (Spelling out the Markov chain requirement we have a distribution over $(v,t,k,s)$ such that the choice of $t$ is conditioned only on $v$, the choice of $k$ is conditioned only on $t$, and finally the choice of $s$ is conditioned only on $k$.)

We say that \((V,T,K,S)\) are a \((\lambda,\delta,\alpha)\) -\emph{MAS} if
\begin{enumerate}
    \item There is an agreement expander \(\A\) with vertex set \(S\) and edge labels \(K\) so that:
    \begin{itemize}
        \item The marginal distribution of sampling a labeled edge \(\set{s,s'}_k\) in \(\A\), and returning \(s,k\), is the same as the marginal distribution \(s,k\) of the Markov chain.
        \item \(\A\) is \((K,\alpha)\)-sound for \(\delta\)-ensembles. 
    \end{itemize}
    \item The bipartite containment graph of \(K\) vs. \(T\), is a \(\lambda\)-sampler. Here the probability of sampling an edge \((k,t)\) is the probability of sampling \((k,t)\) together in the Markov chain.
\end{enumerate}
\end{definition}

A natural example for an MAS is the Grassmannian complex, that is, a four layer structure where \(V = \mathbb{F}_p^n\) and \(T,K,S\) are affine spaces of \(\mathbb{F}_p^n\) of fixed dimensions. We elaborate on this example in the subsection below.
The Grassmannian complex is dense, that is, the number of subspaces grows exponentially with the dimension. No known codes on the Grassmannian complex have good rate. 

Currently known constant degree MASs arise from high-dimensional expanders which are simplicial complexes. However, we cannot use MASs that are directly simplicial complexes to construct any code with non-trivial rate and distance. It is conceivable that high-dimensional expanders that are not simplicial complexes\footnote{These can still arise from high dimensional expanders. For example an MAS whose subsets are {\em links} of a high dimensional expander.} may yield good LTCs.

\subsection{MASs coming from the Grassmannian Complex}
The set system for the Grassmannian MAS is corresponds to points and affine subspaces of a vector space. Formally, let \(p\) be some prime power, and \(q_0<q_1<q_2<n\) be some integers greater than \(0\). Our ground set is \(V = \mathbb{F}_p^n\), and over it we define the following set system \(X=(V,T,K,S)\) where \(T,K,S\) consist of all affine subspaces of dimensions \(q_0,q_1\) and \(q_2\) respectively. The Markov process of this set system, is sampling \((v \in t \subseteq k \subseteq s)\) uniformly.

The edge distribution of the test graph \(\set{s,s'}_k \sim \A\), is to sample a subspace \(k \in K\), and then two subspaces \(s,s'\in S\) independently, given that \(s,s'\supset k\). We call this the \(q_2,q_1\)-agreement test.

We claim that this set system is an MAS:

\begin{lemma}
\label{lem:Grassmann-is-MAS}
There is a universal constant \(\alpha > 0\) so that the following holds. Let \(q_0<q_1<q_2<n\) be as above, and assume \(q_2 \geq 3q_1+2\). Let \(p\) be any prime power.
Let \(X=(V,T,K,S)\) be as above. Then \(X\) is a \((p^{q_0-q_1},\delta, \delta \alpha)\)-MAS for every \(\delta > 0\).
\end{lemma}
the constant above does not depend on \(p\), nor on \(q_0,q_1,q_2,n\).

\begin{proof}
The sampling properties of the layers of a Grassmannian complex are folklore:
\begin{fact}\torestate{ \label{fact:grassmann-samplers}
Let \(G=(K,T,E)\) be the graph where \(K\) are subspaces \(\mathbb{F}_p^n\) of dimension \(q_1\) and \(T\)  are subspaces of dimension \(q_0\), and \((t,k) \in E\) if \(t \subset k\) with uniform weights. This graph is a \(p^{-|q_0-q_1|}\)-sampler. }
\end{fact}

Agreement of the \(q_2,q_1\)-agreement test graph was proven by \cite{DiksteinD2019} (Theorem 6.2).
\begin{theorem}[Agreement for Grassmannian]
\torestate{\label{thm:agreement-on-Grassmann-affine}
There exists a constant $\alpha > 0$ such that for every prime power $p$, $\delta > 0$, and integers $q_1,q_2,n$ such that $3q_1+2 < q_2 \leq n$ the following holds. The $q_2,q_1$-Grassmannian agreement test is $(K,\delta \alpha)$-sound for $\delta$-ensembles.}
\end{theorem}

Combining these two statements together we get that there exists some \(\alpha>0\) so that for every \(\delta>0\), \((V,T,K,S)\) defined above are a \((p^{q_0-q_1},\delta,\delta\alpha)\)-MAS.
\end{proof}

In \pref{sec:applications} use our main theorem, \pref{thm:LLTC-imples-GLTC}, to show that local testability of lifted on the Grassmannian complex, is implied by the local testability of the base code.

\section{Main Theorem - Locally Testable Codes on MASs}
Given an \(MAS\) \((V,T,K,S)\) and a set of base codes \(\sett{C_t}{t\in T}\), the \emph{lifted code} to \(V\) is
\[C = \sett{w:V\to \Sigma}{\rest{w}{t}\in C_t, \forall t\in T}.\]
Similarly, for every \(s \in S\) or \(k \in K\), the local lifts to \(s\) or \(k\) are
\[C_s = \sett{w:s\to \Sigma}{\rest{w}{t}\in C_t, \forall t\subseteq s}, \;
C_k = \sett{w:k\to \Sigma}{\rest{w}{t}\in C_t, \forall t\subseteq k}.\]

The next theorem is a reformulation of \pref{thm:main-hdx}.
\begin{theorem}[Main] \label{thm:LLTC-imples-GLTC}
Let \(V\) be a finite set and \(\rho,\delta,\lambda,\alpha \geq 0\) so that \(\lambda \leq  \frac{\rho \delta \alpha}{64} \). Let \(X = (V,T,K,S)\) be a \((\delta,\lambda,\alpha)\)-MAS. Let \(\sett{C_t}{t\in T}\) be a set of base codes, and let \(C\) be the lifted code. Suppose that
\begin{enumerate}
    \item \emph{Local Distance:} \(C_k\) has distance \(\delta\) for every \(k \in K\).
    \item \emph{Local local testability:} For every \(s \in S\), the code \(C_s\) is \(\rho\)-locally testable with respect to sampling \(t \in T\) given that \(t \subset s\).
\end{enumerate}
Then \(C\) is \(\frac{\rho \delta \alpha}{16}\)-locally testable with respect to the distribution of choosing \(t \in T\).
\end{theorem}
We encourage the readers to think of \(\lambda,\alpha\) as some fixed constants of the set system. Then the theorem states that if \(\set{C_k}\) have large relative distance \(\delta = \Omega(1)\), and \(\set{C_s}\) are \(\rho\)-locally testable for a large enough \(\rho\), then the lifted code is \(\Omega (\rho)\)-locally testable.

\subsection{Proof of the Main Theorem}
\begin{proof}[Proof of \pref{thm:LLTC-imples-GLTC}]

Let \(w_0: V \to \Sigma\) be some word so that 
\[Fail(w_0) \defeq \Prob[t \in T]{\rest{w_0}{t} \notin C_t} = \varepsilon.\]
We need to find a word \(w^*\) so that \(\dist(w_0,w^*) \leq \frac{16\varepsilon}{\rho \delta \alpha}\). We will find a word \(w_1:V \to \Sigma\) so that \(\dist(w_0,w_1) = \frac{8 \varepsilon}{\rho \delta \alpha}\), and 
\[ Fail(w_1) = \Prob[t \in T]{\rest{w_1}{t} \notin C_t} \leq \frac{1}{2}\varepsilon.\]

As a first step we define a function ensemble \(\sett{f_s}{s \in S}\) so that \(f_s \in C_s\) is the closest code word to \(\rest{w_0}{s}\) (ties broken arbitrarily).
For each $k\subset s,s'$ both $f_s|_k\in C_k$ and $f_{s'}|_k\in C_k$, and since $C_k$ is a code with relative distance $\delta$, we get that $\set{f_s}$ is a \(\delta\)-ensemble.

We claim that the ensemble passes the agreement test with high probability.
\begin{claim}
\label{claim:ensemble-has-high-agreement}
\[\Prob[\set{s_1,s_2}_k \sim \A]{\rest{f_{s_1}}{k} = \rest{f_{s_2}}{k}} = 1- \frac{4 \varepsilon}{\rho \delta}.\]
\end{claim}

As there is an agreement expander \(\A\) that is \((K,\alpha)\)-sound with respect to \(\delta\)-ensembles, there exists some function \(w_1:V \to \Sigma\) so that 
\begin{equation}
\label{eq:agreement-guarantee}
\Prob[k \subset s]{\rest{w_1}{k} = \rest{f_s}{k}} = 1-\frac{4 \varepsilon}{\rho \delta \alpha}.
\end{equation}
We claim that \(w_0\) is close to \(w_1\), and that \(w_1\) fails the test with probability \(\leq \frac{\varepsilon}{2}\).
\begin{claim}
\label{claim:new-is-close-to-old}
\(\dist (w_0,w_1) \leq \frac{8  \varepsilon}{\rho \delta \alpha}.\)
\end{claim}

\begin{claim}
\label{claim:new-function-rarely-fails}
\(Fail(w_1) \leq \frac{1}{2}\varepsilon.\)
\end{claim}

Modulo \pref{claim:new-is-close-to-old} and \pref{claim:new-function-rarely-fails}, we repeat the correction process \(poly(\log (\min_{t \in T} \prob{t}))\) times. In the beginning of the \(i\)-th iteration we start with \(w_i\) that fails the test with probability \( \leq \varepsilon/2^i\). In the end of the iteration, we find \(w_{i+1}\) that fails the test with probability \( \leq \varepsilon/2^{i+1}\), and so that \(\dist(w_i,w_{i+1})\leq \frac{8 \varepsilon}{ \rho \delta \alpha 2^{i}}\). 
Thus we obtain a sequence of functions  \(w_0,w_1,w_2,..., w_r\) that ends with \(w_r = w^*\) that always passes the test. The distance we accumulate from \(w_0\) is
\[ \dist (w_0,w_r) \leq \sum_{i=0}^{r-1}\dist(w_i, w_{i+1}) \leq \frac{8}{\rho \delta \alpha } \sum_{i=0}^\infty \frac{1}{2^i} = \frac{16}{\rho \delta \alpha}.
\]
\end{proof}

\begin{proof}[Proof of \pref{claim:ensemble-has-high-agreement}]

By the local testability of the base code $C_s$,
\begin{equation} \label{eq:small-S-local-distance}
    \Ex[s]{\dist (\rest{w_0}{s},C_s)} \leq \rho^{-1} \Ex[s]{\Prob[t \subset s]{\rest{w_0}{t} \notin C_t}} = \rho^{-1} \Prob[t]{\rest{w_0}{t} \notin C_t} \leq  \frac{\varepsilon}{\rho} .
\end{equation} 
As \(f_s\) is closest code word to \(\rest{w_0}{s}\),
\[ \frac{\varepsilon}{\rho} \geq \Ex[s]{\dist (\rest{w_0}{s},f_s)} = \Ex[s]{\Ex[k \subset s]{\dist (\rest{w_0}{k},\rest{f_s}{k})}}.\]
By Markov's inequality, with probability \(1-\frac{4\varepsilon}{\rho \delta}\) of sampling \(\set{s_1,s_2}_k \sim \A\), it holds that \(\dist(\rest{w_0}{k},\rest{f_{s_i}}{k}) < \frac{\delta}{2}\) where \(f_{s_i}\) is the closest codeword in \(C_{s_i}\) to \(\rest{w_0}{s_i}\). 

By the local distance assumption, \(C_k\) has distance \(\delta\), and if \(\dist(\rest{f_{s_1}}{k},\rest{f_{s_2}}{k}) < \delta\), then
\[\rest{f_{s_1}}{k} = \rest{f_{s_2}}{k}.\]
\end{proof}

\begin{proof}[Proof of \pref{claim:new-is-close-to-old}]
We note that
\[\dist(w_0,w_1) = \Ex[s]{\dist(\rest{w_0}{s},\rest{w_1}{s})}.\]
We show closeness by the triangle inequality. Fix \(s \in S\), then
\[\dist(\rest{w_0}{s},\rest{w_1}{s}) \leq \dist(\rest{w_0}{s},f_s) + \dist(f_s,\rest{w_1}{s}).\]

By \eqref{eq:small-S-local-distance},
\[\Ex[s]{\dist (\rest{w_0}{s},f_s)} \leq \frac{\varepsilon}{\rho}.\]

By the \((K,\alpha)\)-soundness of the agreement expander \(\A\),
\[ \dist(\rest{w_1}{s},f_s) = \Ex[k \subset  s]{\dist(\rest{w_1}{k},\rest{f_s}{k})} \leq \Prob[k \subset s]{\rest{w_1}{k} \ne \rest{f_s}{k}} = \frac{4 \varepsilon}{\rho \delta \alpha}.\]

By the triangle inequality, and using the fact that both \(\delta,\alpha < 1\)
\[\dist (w_0,w_1) \leq \frac{8 \varepsilon }{\rho \delta \alpha}.\]
\end{proof}

The proof of \pref{claim:new-function-rarely-fails} relies on the \(\lambda\)-sampling property of the MAS. 

\begin{proof}[Proof of \pref{claim:new-function-rarely-fails}]
By assumption the containment graph between \(T\) and \(K\) is has the \(\lambda\)-sampling property.
Let 
\(B = \sett{k \in K}{\forall s \supset k, \; \rest{f_s}{k} \ne \rest{w_1}{k} }\).
We observe the following:
\begin{enumerate}
    \item By the agreement property, \(\prob{B} \leq \frac{8\varepsilon }{\rho \delta \alpha}\), and without loss of generality \(\prob{B} \leq \frac{1}{2}\) (if we want to show that the code is \(\frac{\rho \delta \alpha }{16}\)-locally testable, it is enough to consider \(\varepsilon\) so that \(\frac{16 \varepsilon }{\rho \delta \alpha} \leq 1\)).
    \item If \(t \in T\) contributes to the failure (i.e \(\rest{w_1}{t} \notin C_t\)), then \(\rest{w_1}{k} \ne \rest{f_s}{k}\) for all \(k\supset t\) and \(s \supset k\). Thus \emph{all} its neighbours are in \(B\).
\end{enumerate}
Denote by \(N\) the set of \(t \in T\) so that all of \(t\)'s neighbours are in \(B\). By item \(2\) above we have that \(\Prob[t \in T]{\rest{w_1}{t} \notin C_t} \leq \Prob[t \in T]{N}\). We note that if we sample a neighbour of \(t\), we get some \(k \in B\) with probability \(1\geq \prob{B} + \frac{1}{2}\).
Thus by the \(\lambda\)-sampling property, we have that
\[ \Prob[t \in T]{N} \leq 4\lambda \frac{8 \varepsilon}{\rho \delta \alpha}.\]
We chose \(\lambda \leq \frac{\rho \delta \alpha}{64}\), hence \(Fail(w_1) \leq \frac{1}{2}\varepsilon\).

\end{proof}

\begin{remark}
The MAS has four layers. The vertex layer \(V\) and the layer \(T\) are required to define the lifted code itself. It is also natural to introduce a higher layer \(S\), since without any other requirements we can't expect any lifted code to be locally testable.

However, the intermediate layer \(K\) is possibly unneeded. While it is a crucial part of the \emph{proof}, it is not needed for lifting the code, nor for the local tests. We believe it is interesting to understand whether it is enough to study a three-layered set system, namely \((V,T,S)\). Are there similar properties, in terms of agreement, sampling and expansion, that also give us a similar result?
\end{remark}

\section{Local Testability in Vector Spaces} \label{sec:applications}

In this section we demonstrate how the main theorem fits in with, and generalizes, the known results on testability of Reed-Muller codes. In this case the MAS is the Grassmannian complex MAS described in \pref{lem:Grassmann-is-MAS} for \(V=\mathbb{F}_p^n\) and \(T,K,S\) being the collections of all affine subspaces of dimension \(q_0,q_1,q_2\) respectively.

We define the code on \(V\) by lifting base codes \(\sett{C_t}{t \in T}\). Namely
\[ C = \sett{w:\mathbb{F}_p^n\to\mathbb{F}_p }{\rest{w}{t} \in C_t, \; \forall t\in T }. \]

One example of such a code, is the \((n,r)\)-Reed-Muller code on \(\mathbb{F}_p^n\). This code consists of all polynomials of degree \(\leq r\). When \(n=1\) we call this the Reed-Solomon code. Take \(T\) to be the set of all affine lines (i.e. \(q_0=1\)), and let \(C_t\) be the \(r\)-Reed-Solomon code on every line. Lifting this code to \(V\) results in all functions \(w:\mathbb{F}_p^n \to \mathbb{F}_p\) so that for every line \(t \in T\), \(\rest{f}{t}\) is a function of degree at most \(r\). For some parameters \(n,r,p\) this results in the \((n,r)\)-Reed-Muller code. Surprisingly, \cite{GuoKS2013} showed that for some other parameters \(r,n,p\) the code lifted from the \(r\)-Reed-Solomon code, contains more than the \((n,r)\)-Reed-Muller code. Nevertheless, these codes are locally testable as well \cite{GuoHS2015,HaramatyRS2015}.

Our main theorem states that to prove local testability of \(C\) it is enough to prove that \(C_s = \sett{w:s \to \mathbb{F}_p }{\rest{w}{t} \in C_t, \; \forall t\in T, t\subset s } \) is locally testable, for to each subspace \(s \in S\). 

This gives rise to testability results for Reed-Muller codes (which are well studied, see \cite{RubinfeldS1996, RazS1997, AroraS2003}) as well as to lifted codes as were studied in  \cite{GuoKS2013} (given of course, that we check their local testability in a some small fixed space). Moreover, this statement continues to hold for more general sets of base codes \(\sett{C_t}{t \in T}\):  If the lifts of \(\sett{C_t}{t\in T}\) to dimension \(q_2\) subspaces are locally testable (with good enough parameters), then the lifted code to dimension \(n\) is also locally testable. This is particularly useful in the regime where \(q_0,q_2\) are fixed, and \(n\) tends to infinity. This includes the examples above, but is a more general statement.

\begin{theorem}
\label{thm:tanner-grassmann}
There is a universal constant \(\alpha>0\) so that the following holds. Let \(q_0<q_1<q_2<n\) be as above, and assume \(q_2 \geq 3q_1+2\). Let \(p\) be any prime power.
Let \(X=(V,T,K,S)\) be as above.
Let \(\sett{C_t}{t\in T}\) be a set of base codes, and suppose that there exists
some \(\delta > 0\) and \(\rho \geq \frac{64 p^{q_1-q_0}}{\alpha \delta^2}\) so that:
\begin{enumerate}
    \item For any \(q_1\) dimensional space \(k \in K\), \(C_k\) has distance \(\geq \delta\).\footnote{\cite{GuoKS2013} showed this holds, for example, whenever the base codes \(C_t\) themselves have distance \(\geq \delta + \frac{1}{p^{q_0}}\).} 
    \item For every \(q_2\) dimensional space \(s \in S\), \(C_s\)
    is \(\rho\)-locally testable.
\end{enumerate}
Then for any \(n > q_2\), the lift of \(\sett{C_t}{t\in T}\) to \(\mathbb{F}_p^n\) is \(\frac{\rho \delta^2 \alpha}{16}\)-locally testable.
\end{theorem}
The constant \(\alpha\) doesn't depend on any of the other parameters, nor on the field size.

We encourage the readers to think of \(\delta = \Omega(1)\). Then for every fixed dimensions \(q_0,q_1\) and field size \(p\) there is some \(\rho\), so that for every lifted code that is \(\rho\)-locally testable on spaces of dimension \(q_2\), the code is also \(\Omega(\rho)\)-locally testable on all spaces of dimension \(n>d\) (for a large enough \(\rho\)). Note that this theorem applies both to the regime where the field size is small (e.g. \(p=2,3\)), and where the field size goes to infinity. When \(p\) grows, the conditions of the theorem become easier to satisfy, that is, that the lower bound on \(\rho\) becomes smaller as well.

\begin{proof}[Proof of \pref{thm:tanner-grassmann}]
Let \(\alpha\) be the constant stated in \pref{lem:Grassmann-is-MAS}. The system \(X=(V,T,K,S)\) defined above is a \((p^{q_0-q_1},\delta, \delta \alpha)\)-MAS, by \pref{lem:Grassmann-is-MAS}, for that \(\alpha\).

Denote by \(C\) the lift of \(\sett{C_t}{t \in T}\) to \(\mathbb{F}_p^n\). This code satisfies the distance and local local testability properties:
\begin{enumerate}
    \item The lift of \(\sett{C_t}{t\in T}\) to an \(q_1\) dimensional space \(k \in K\) has distance \(\geq \delta\).
    \item The lift of \(\sett{C_t}{t\in T}\) to a \(q_2\)-dimensional space \(s \in S\) is \(\rho\)-locally testable.
\end{enumerate}
Hence by \pref{thm:LLTC-imples-GLTC}, this code is \(\frac{\rho \delta^2 \alpha}{16}\)-locally testable.
\end{proof}

%\printbibliography
\renewcommand{\baselinestretch}{1.0}

{\small
\bibliographystyle{prahladhurl}

\bibliography{ltchdx-bib}
}

%\appendix
%\input{appendix.tex}
\end{document}